\documentclass[reprint, aps, prd, letterpaper, nofootinbib, superscriptaddress, longbibliography, notitlepage]{revtex4-2}
\usepackage[utf8]{inputenc}
\usepackage{amsmath,amssymb,amsfonts}
\usepackage{mathrsfs}
\usepackage{color}
\usepackage{graphicx} 
\usepackage[section]{placeins}
\usepackage[colorlinks=true,linkcolor=blue,citecolor=blue]{hyperref}
\usepackage{comment}

\allowdisplaybreaks

\pdfoutput=1 

\usepackage{graphicx}
\usepackage{hyperref}
\usepackage{aas_macros}
\usepackage{amsmath}
\usepackage{amssymb}
\usepackage{xcolor}
\usepackage{nccmath}
\usepackage{cleveref}

\newcommand{\GeV}{~\text{GeV}}

\newcommand{\Msol}{~M_{\odot}}

\newcommand{\cms}{\text{cm}~\text{s}^{-1}}
\newcommand{\kpc}{\text{kpc}}

\def\beq{\begin{equation}\begin{aligned}}
\def\eeq{\end{aligned}\end{equation}}

\begin{document}

\title{\boldmath Structure formation with
 dark magnetohydrodynamics}

\author{Pierce Giffin}
\email{pgiffin@ucsc.edu}
\affiliation{Department of Physics, 1156 High Street, University of California Santa Cruz, Santa Cruz, California 95064, USA}
\affiliation{Santa Cruz Institute for Particle Physics, 1156 High Street, Santa Cruz, California 95064, USA}

\author{Andrew Liu}
\email{al1279@alumni.princeton.edu}
\affiliation{Department of Physics, Princeton University, Princeton, New Jersey 08544, USA}

\author{Jeremias Boucsein}
\email{pv276@uni-heidelberg.de}
\affiliation{Department of Physics, 1156 High Street, University of California Santa Cruz, Santa Cruz, California 95064, USA}
\affiliation{Universität Heidelberg, Zentrum für Astronomie, Institut für theoretische Astrophysik, Albert-Ueberle-Strasse 3, 69120 Heidelberg,
Germany}

\author{Akaxia Cruz}
\email{akaxia@princeton.edu}
\affiliation{Center for Computational Astrophysics, Flatiron Institute, New York, New York 10010, USA}
\affiliation{Department of Physics, Princeton University, Princeton, New Jersey 08544, USA}
\affiliation{Department of Astrophysical Sciences, Princeton University, Princeton, New Jersey 08544, USA}

\author{Anirudh Prabhu}
\email{prabhu@princeton.edu}
\affiliation{Department of Physics, Princeton University, Princeton, NJ 08544, USA}
\affiliation{Princeton Center for Theoretical Science, Princeton University, Princeton, NJ 08544, USA}

\author{Stefano Profumo}
\email{profumo@ucsc.edu}
\affiliation{Department of Physics, 1156 High Street, University of California Santa Cruz, Santa Cruz, California 95064, USA}
\affiliation{Santa Cruz Institute for Particle Physics, 1156 High Street, Santa Cruz, California 95064, USA}

\author{M. Grant Roberts}
\email{migrober@ucsc.edu}
\affiliation{Department of Physics, 1156 High Street, University of California Santa Cruz, Santa Cruz, California 95064, USA}
\affiliation{Santa Cruz Institute for Particle Physics, 1156 High Street, Santa Cruz, California 95064, USA}

\begin{abstract}
\noindent Long-range interactions in the dark sector can give rise to collective plasma phenomena that are capable of modifying the evolution of dark matter halos. We present the first study of gravitational collapse in a secluded dark $U(1)_D$ model using a magnetohydrodynamic description of the dark matter. We show that dark magnetic fields generate an anisotropic pressure that alters the Jeans scale and suppresses small-scale power in a direction-dependent manner. For a range of primordial magnetic spectral indices, this effect produces distinctive modifications to the linear matter power spectrum. We find that current observations cannot yet constrain viable dark magnetic fields, as cosmic microwave background (CMB) tensor modes mostly provide more stringent constraints. Nevertheless, forthcoming high-resolution probes of the matter power spectrum (CMB-HD lensing, HERA, and EDGES) will be able to test these predictions and are sensitive to dark charge-to-mass ratios in the range $10^{-20}\,\text{GeV}^{-1}\lesssim q_\chi/m_\chi\lesssim 10^{-14}\,\text{GeV}^{-1}$. 
\end{abstract}

\maketitle
\newpage
\section{Introduction}
\label{sec:intro}

The standard cosmological paradigm, the $\Lambda$CDM model, predicts that the matter content in the Universe is dominated by cold dark matter (CDM), a cold, collisionless form of matter that interacts with itself and with baryonic matter only gravitationally. The existence of CDM is supported by numerous observations at large scales such as the cosmic microwave background (CMB)~\cite{Planck:2018vyg} and large-scale structure of the Universe~\cite{Springel:2006vs}. However, a much broader range of dark matter (DM) models are capable of reproducing the large-scale successes of CDM, while producing different observables at smaller scales. For an up-to-date review of various DM paradigms see~\cite{PDG, profumo_introduction_2017}. A minimal extension to the CDM paradigm is to allow for nongravitational self-interactions between DM particles. This self-interacting DM (SIDM) was originally introduced to solve galactic scale anomalies~\cite{Spergel:1999mh}, and is currently used to alleviate small structure issues, such as the core-cusp problem, diversity of rotation curves, and too-big-too-fail \cite{de_Blok_2009, Oman_2015, Kaplinghat:2019dhn, Ren:2018jpt, Relatores:2019ews, Zentner:2022xux, Roberts2025, Boylan_Kolchin_2011}, and is also well motivated from a particle physics perspective (see, e.g.,~\cite{Tulin:2017ara,Adhikari:2022sbh} for  recent reviews).  


A majority of work in SIDM has focused on the scenario in which DM particles interact through a short-range force mediated by a boson with mass much greater than the typical momentum transfer, $q \sim v_{\rm DM} m_{\rm DM}$, where $v_{\rm DM}$ and $m_{\rm DM}$ are the typical velocity and mass of DM particles. In this limit, particles only interact via individual hard scattering events. Recently, there has been growing interest in the case in which DM possesses coherent, long-range self-interactions~\cite {Bogorad:2023wzn}. A subset of these models consider the scenario in which the dark force is mediated by an ultralight vector~\cite{Ackerman_2009, Lasenby:2020, Cruz_2023, Giffin, Medvedev:2024kjh, Khlopov_1, Khlopov_2, Khlopov_3, Ackerman:2008kmp, Agrawal:2016quu, Cyr-Racine:2012tfp}. If the particle collision rate is sufficiently small, these long-range interactions are dominated by collective effects. For instance, DM may be susceptible to velocity-space instabilities that lead to resonant exchange of energy between particles and (mediator) waves and can dramatically alter DM kinematics (see, e.g.,~\cite{Bhattacharjee2017} for a review of such processes in Standard Model plasmas).


In such scenarios, formation of plasma instabilities may occur in the dark sector, dramatically altering the system's dynamics. The possibility of dark-sector plasma instabilities has been explored in several contexts \cite{Ackerman_2009, Lasenby:2020, Cruz_2023}. In the visible sector, analyses of linear instabilities in environments such as supernova remnants, the CMB, and merging galaxy clusters \cite{Li_2020, Cruz_2023} have yielded stringent constraints on models where dark-sector particles interact with Standard Model (SM) electrons, since electron–electron collisions efficiently quench instability growth. In contrast, constraints on secluded dark sectors, where such collisional damping is absent, are considerably weaker. In these models, both electromagnetic and electrostatic modes in the dark plasma can grow largely unimpeded, allowing instabilities to persist over cosmological timescales. This broader instability landscape motivates a detailed study of their impact on structure formation and on astrophysical observables.

Beyond linear theory, numerical simulations have begun to explore the nonlinear dynamics of dark-sector plasmas. Hydrodynamic simulations of cluster collisions under a two-component DM model, treating part of the DM as a fluidlike plasma, have been able to reproduce density maps that closely resemble Bullet Cluster and Abell 520 in the case where a significant fraction of the DM behaves as a plasma \cite{Heikinheimo_2015, Spethmann_2017, Heikinheimo_2018}. More recently, fully kinetic simulations of counter-streaming dark-matter flows, combined with observations of the Bullet Cluster, have further constrained the parameter space of secluded dark $U(1)_D$~\cite{Giffin}. These studies highlight the potentially significant astrophysical consequences of plasmalike interactions in the dark sector.

However, one critical ingredient remains largely unexplored: the role of gravity in the formation and evolution of dark-sector plasma instabilities. Understanding its interplay with dark plasma dynamics is essential for determining the impact on both primordial density perturbations and late-time structure formation. In this work, we incorporate gravitational effects into the analysis of dark-sector plasma instabilities for the first time. We characterize the alterations to primordial structure formation resulting from the formation of dark plasma instabilities. The paper will be structured as follows: 
\begin{enumerate}
    \item We begin in Sec. \ref{sec_2} with a brief review of the classical gravitational instability without any alterations from dark sector plasma effects. This is a primer for the work done in Sec. \ref{sec_3}. 
    \item In Sec. \ref{sec_3} we introduce the dark $U(1)_D$ model that we consider throughout this work, and couple dark sector plasma effects to the classical Jeans instability, demonstrating a directionally dependent alteration to both the Jeans length and group velocity in the presence of an external dark magnetic field.
    \item Section \ref{sec_4} explores the possible generation mechanisms of background dark sector magnetic fields. We introduce the primordial dark magnetic field power spectra and provide conservative constraints on their present-day amplitudes. 
    \item We then present the alterations to the linear matter power spectrum under this model in Sec. \ref{sec_5}. Scanning over parameter space and performing a $\chi^{2}$ analysis against current and future measurements of the primordial matter power spectrum, we also show our constraints on the background dark magnetic field, as well as the charge-to-mass ratio for this model for various different spectral indices for the primordial magnetic field spectrum.
    \item Finally, with Sec. \ref{sec_6} we conclude the paper with a discussion of not only the constraints we derived, but also possible future extensions of our work, such as the consideration of kinetic mixing with the Standard Model photon and the possibility of halo triaxiality as a result of anisotropic collapse due to the directionally dependent alteration to the Jeans length.
\end{enumerate}

\section{Gravitational Collapse and classical Jeans length} \label{sec_2}

The classical Jeans length is given by $\lambda_{J} = \frac{2 \pi}{k_J}$, where $k_J$ is given by 
\begin{equation}
    k_J=\left(\frac{4\pi G \rho_0}{c_s^2}\right)^{1/2}
\end{equation}
Here, $c_s$ denotes the speed of sound in the medium, and $\rho_0$ is the unperturbed mass density of the ensemble. This defines the critical length at which the gas becomes stable against gravitational collapse. Specifically, only perturbations with wavelengths that exceed the Jeans length $\lambda > \lambda_J$ will trigger instabilities \cite{binney_galactic_2008}. 

While the traditional Jeans length formula $\lambda_J =\sqrt{\frac{\pi c_s^2}{G\rho}}$ breaks down for CDM due to its zero sound speed \cite{Weinberg:2008zzc}, an effective Jeans length can be defined that varies with scale. At small scales, this effective Jeans length is negligibly small, allowing CDM to cluster on all scales. On larger scales, the expansion of the Universe acts as an effective pressure, leading to a comoving Jeans length approximated by $\lambda_J  \approx \frac{2\pi c_s}{H a \sqrt{\Omega_m}}$, where $H$ is the Hubble parameter, $a$ is the scale factor, and $\Omega_m$ is the matter density parameter \cite{Barkana_2001}. The effective sound speed $c_s$ is related to the velocity dispersion $\sigma(R)$ at a given scale $R$ by $c_s^2 \approx \frac{\sigma^2(R)}{3}$. The velocity dispersion itself scales with the enclosed mass as $\sigma_{200} = A [h(z) \frac{M_{200}}{10^{15} M_\odot}]^\alpha$, where $h(z)$ is the dimensionless Hubble parameter, $A \approx 1000$ km/s, $\alpha \approx 0.35-0.36$ \cite{Evrard:2007py, Munari_2013}. This leads to a scale-dependent Jeans length that scales approximately linearly with $R$: $\lambda_J(R) \propto R$. This scale-dependent Jeans length is crucial for understanding structure formation in CDM models, supporting the bottom-up scenario of hierarchical clustering. Unlike warm or hot DM, CDM lacks a definitive cutoff scale for structure formation, which contributes to the small-scale challenges in $\Lambda$CDM cosmology \cite{Bullock:2017xww}.

\section{Evolution of Dark Magnetized Plasma} \label{sec_3}
\label{sec:DarkPlasmas}
To allow for collective plasma effects, we consider a dark sector consisting of dark positrons and electrons of equal abundance that interact through a massless $U(1)_D$ gauge boson, which we will refer to as a ``dark photon." The dark sector Lagrangian for this model takes the form
\begin{equation}
    \mathcal{L}_{\rm Dark} = -\frac{1}{4}F'^{\mu \nu}F'_{\mu\nu} + \bar{\chi}(\gamma^\mu(i\partial_{\mu}-q_{\chi}A'_\mu)-m_\chi)\chi 
\end{equation}
where $\chi$, $\bar{\chi}$ are dark electrons and positrons, respectively, with charge $q_\chi$ and mass $m_\chi$, and $A'$ is the dark photon. An additional term of the form $
\frac{\epsilon}{2}F'^{\mu \nu}F_{\mu \nu}$ would allow for the possibility of kinetic mixing between the dark photon and the Standard Model photon. In this work, we will only consider the case where kinetic mixing does not occur, giving rise to a secluded dark sector. Note that if $\epsilon$ is set to zero at some high energy scale, it will remain zero under renormalization group evolution unless there are heavy states charged under both $U(1)_D$ and $U(1)_{\rm EM}$~\cite{Ackerman:2008kmp}. We leave the study of $\epsilon\neq0$ for future work and restrict the model to the scenario of a massless dark photon. 

Previous works have utilized the evolution of beam instabilities in plasmas to derive constraints on such a model \cite{Cruz_2023, Li_2020, Giffin}. For the purpose of characterizing the effects of plasma instabilities and how they alter structure formation in the Universe, beam-type instabilities do not provide the correct picture for this scenario. Instead, we consider a homogeneous and isotropic, net-neutral plasma of particles to characterize the criteria for gravitational collapse in the presence of collective plasma effects. 

The evolution of a plasma can be well-approximated by a fluid model. Though this assumption is typically made in highly collisional systems, it can also provide a framework to study the evolution of collisionless systems whose velocity distributions remain approximately Maxwellian. In a collisional plasma, particle collisions drive the distribution function toward a Maxwellian, as explained by Boltzmann’s H-theorem. Even in collisionless plasmas, collective kinetic processes such as phase mixing, Landau damping, wave-particle scattering, and microinstabilities can produce distributions that are close to Maxwellian over macroscopic scales, despite the absence of true collisions. These systems may, however, evolve away from Maxwellians (e.g., toward power-laws) on longer timescales~\cite{Ewart:2024bnh}. We leave a more detailed treatment of these effects to future work. Treating the plasma as a fluid yields the system of partial differential equations often referred to as the magnetohydrodynamic (MHD) equations. Adopting a system of units in which $\hbar=c=\epsilon_0=\mu_0=1$, the system of equations is
\begin{align}
    &\nabla\cdot \vec{E} = \rho_c & \nabla \times \vec{E} = -\frac{\partial \vec{B}}{\partial t}\\
    & \nabla \cdot \vec{B} = 0 & \nabla \times \vec{B} =  \vec{J}+ \frac{\partial\vec{E}}{\partial t}\\
    &\frac{\partial \rho_m}{\partial t}+\nabla \cdot (\rho_m \vec{U})=0 & \rho_m \frac{d\vec{U}}{dt}=\vec{J}\times \vec{B} -\nabla\cdot \stackrel{\leftrightarrow}{P}\\
    & \frac{d}{dt}\left(\frac{p_\perp B^2}{\rho_m^3}\right)=0 & \frac{d}{dt}\left(\frac{p_\parallel}{\rho_m B}\right)=0\\
    &\vec{E}+\vec{U}\times\vec{B}=\frac12\frac{m_\chi^2}{q_\chi^2\rho_m}\frac{\partial\vec J}{\partial t}
\end{align}
Here $\vec{U}$ is the bulk velocity of the DM fluid,  $\rho_c$ and $\rho_m$ are the charge and mass density of the fluid, respectively. $\stackrel{\leftrightarrow}{P}$ is the pressure tensor containing only diagonal elements of $p_\perp$ in the directions perpendicular to the magnetic field and $p_\parallel$ in the direction parallel to the magnetic field. Additionally, we have assumed the Chew-Goldberger-Low (CGL) equation of state for cylindrically symmetric magnetized plasmas \cite{CGL:1956}. The last equation is a specific scenario of the generalized Ohm's law discussed in Appendix \ref{ap:Ohms}.

To incorporate gravitational effects into our system of equations, we only need to include Poisson's equation,
\begin{equation}
    \nabla^2V=4\pi G\rho_m,
\end{equation}
where $V$ is the gravitational potential, along with a correction to the conservation of momentum equation
\begin{equation}
     \rho_m \frac{d\vec U}{dt}=\vec J\times \vec B -\nabla\cdot \stackrel{\leftrightarrow}{P}-\rho_m\nabla V.
\end{equation}

We consider the case of an infinitely large homogeneous self-gravitating plasma in the presence of a uniform background magnetic field. After perturbing this initial state,  small perturbations become unstable and trigger the Jeans instability. Considering electromagnetic effects, we find that such perturbations obey the following dispersion relation in the linear regime
\begin{widetext}
\begin{align}
    \label{eq:Disp_Rel}
     \left[\rho\omega^2-(1+\beta)k_\perp^2p-\beta B^2k^2+4\pi G\rho^2\frac{k_\perp^2}{k^2}\right]
    \left(\rho\omega^2-3k_\parallel^2p+4\pi G \rho^2\frac{k_\parallel^2}{k^2}\right) \nonumber\\
    =k_\perp^2k_\parallel^2\left[\left(p-\frac{4\pi G \rho^2}{k^2}\right)^2+2\left(p-\frac{4\pi G \rho^2}{k^2}\right)\left(1-\beta\right)p\right]
\end{align}
\end{widetext}
with
\begin{equation}
    \beta=\frac{1}{1+\frac{k^2}{2\rho}\frac{m_\chi^2}{q_\chi^2}}
\label{eq:chi-equation}
\end{equation}
Here $p$ is the unperturbed initial isotropic pressure. A detailed derivation is given in Appendix \ref{ap:Disp}. It is assumed that all fields are proportional to $\mathrm{exp}\, i(\vec{k}\cdot\vec{r}-\omega t)$ and the perpendicular and parallel directions are measured with respect to the background magnetic field, $B$. Note that we assume $\rho_c$ is negligibly small due to the large time and length scales of the system of interest. Hence, we drop the $m$ subscript from $\rho_m$ for ease of notation.

\subsection{Parallel propagation}
We first consider waves oriented parallel to the magnetic field ($k_\perp=0$); Eq.\eqref{eq:Disp_Rel} then reduces to
\begin{equation}
    \left[\rho\omega^2-k_\parallel^2\beta B^2\right]
    \times\left(\rho\omega^2-3k_\parallel^2 p+4\pi G\rho^2 \frac{k_\parallel^2}{k^2}\right)=0\,.
\end{equation}
This leads to two potential solutions. The first
\begin{equation}
    \frac{\omega^2}{k_\parallel^2}=\frac{\beta B^2}{\rho},
\end{equation}
only contains solutions with real-valued $\omega$. Hence, no unstable modes are present. The second solution
\begin{equation}
    \frac{\omega^2}{k_\parallel^2}=\frac{3p_\parallel}{\rho}-\frac{4\pi G \rho}{k_\parallel^2},
\end{equation}
shows that the Jean's instability can trigger if
\begin{equation}
    k_\parallel<\left(\frac{4\pi G \rho^2}{3p}\right)^{1/2}
\end{equation}
which is independent of $\beta$. This implies that the speed of sound in the parallel direction is unaltered by the background magnetic field.
\begin{equation}
c_{s,\parallel}^2 \;\equiv\; \frac{3\,p}{\rho}.
\label{eq:cspar}
\end{equation}
Using the CGL equation of state, we can rewrite the sound speed in terms of the dispersive velocity of the DM particles,
\begin{align}
    p_{\parallel} &= nk_B T = \rho v_{\rm th}^2,
\end{align}
allowing us to simplify the effective parallel group velocity to the following form
\begin{equation}
    c^2_{s,\parallel} = 3 v_{\rm th}^2 \,.
\label{eq:parallel_sound_speed}
\end{equation}

\subsection{Perpendicular propagation}
For modes perpendicular to the background magnetic field ($k_\parallel=0$), the dispersion relation contains both perpendicular pressure and electromagnetic support,
\begin{equation}
    \left[\rho\omega^2-k_\perp^2\left(p(1+\beta)+\beta B^2\right)+4\pi G \rho^2\right]=0.
\end{equation}
These modes only contain a single solution for $\omega$
\begin{equation}
    \frac{\omega^2}{k_\perp^2}=\frac{p(1+\beta)+\beta B^2}{\rho}-\frac{4\pi G \rho}{k_\perp^2}\,.
\end{equation}
Here, we only have a Jean's-type instability that triggers when
\begin{equation}
    k_\perp<\left(\frac{4\pi G \rho^2}{p(1+\beta)+\beta B^2}\right)^{1/2}~.
\end{equation}
This leads to an effective perpendicular group velocity that receives contributions from the Alfv\'en group velocity, $v_A=B/\sqrt{\rho}$
\begin{equation}
c_{s,\perp}^2 \;\equiv\; \frac{(1+\beta)\,p}{\rho} + \beta\frac{\,B^2}{\rho}.
\label{eq:csperp}
\end{equation}
We can simplify the perpendicular group velocity using the CGL equation of state for the perpendicular component,
\begin{align}
    p_{\perp} &= \frac{\rho v_{\rm th}^2}{2}  ,
\end{align}
giving us
\begin{equation}
    c^2_{s,\perp} = \frac{v_{\rm th}^2}{2} + \beta\left(\frac{v_{\rm th}^2}{2} + \frac{B^2}{ \rho}\right)\,.
\label{eq:perp_sound_speed}
\end{equation}
Note that in the limit of $q_\chi/m_\chi\to0$, we have $\beta\to 0$. This shows that the plasma decouples from the background magnetic field in the small charge limit, and we retain the classic Jeans length for a cylindrically symmetric system. In the following sections, we investigate the modification of the sound speed in motivated magnetic field models.

\section{Dark Magnetic Fields: Generation, Cosmological Constraints, and Caveats} \label{sec_4}
\label{sec:dark_magnetic_constraints}

If DM particles carry charge under this hidden gauge group, the dark plasma can support dark electric and magnetic fields, collective excitations, and MHD phenomena analogous to those in the visible sector. A central open question is whether dark magnetic fields can be dynamically significant during cosmic structure formation. In such scenarios, gravitational collapse competes with magnetic pressure and tension, modifying the DM power spectrum and the internal structure of halos. In the following subsections, we review the evolution of SM cosmological magnetic fields as well as specific constraints that should additionally provide constraints on dark magnetic fields.


\subsection{Dark magnetic field evolution}

Translating constraints at a given cosmological epoch (e.g., big bang nucleosynthesis (BBN) or recombination) to one at structure formation requires understanding how the magnetic field evolves under cosmological evolution. The evolution differs considerably depending on whether the magnetic field is turbulent or ordered. Heuristically, the transition from order to turbulence depends on the magnetic Reynolds number, ${\rm Re}_M \equiv v_\ell \ell/\eta$, where $v_\ell$ is the velocity of the (dark) plasma on scale $\ell$, and $\eta$ is the (dark) magnetic diffusivity, defined as $\eta \equiv 1/\sigma$, where $\sigma$ is the (dark) conductivity. We assume the dark sector is fully ionized at all relevant epochs, which is expected since the dark photon temperature is much larger than the binding energy of dark positronium in the relevant parameter space. The dark conductivity is then

\begin{align}
    \sigma_{\chi} = \frac{q_\chi^2 n_\chi}{m_\chi \nu_{\chi \chi}} = \frac{{12 \pi^{3/2} T_\chi^{3/2}}}{q_\chi^2 m_\chi^{1/2} \ln \Lambda},
\end{align}
where $\nu_{\chi \chi} = {q_\chi^4 n_\chi \ln \Lambda / 12 \pi^{3/2} m_\chi^{1/2} T_\chi^{3/2}}$ is the Coulomb scattering frequency of $\chi$ particles\footnote{Exclusion of $\chi, \bar{\chi}$ collisions does not alter our conclusions.}, $\ln \Lambda$ is the Coulomb logarithm factor, and $T_\chi$ is the DM temperature. As we will be considering very small dark charge, we will typically have ${\rm Re}_M \gg 1$, which means the evolution of magnetic fields is well-described by ideal MHD:

\begin{align}
    \frac{\partial \vec{B}_{\rm{com}}}{\partial t} = \frac{1}{a} \vec{\nabla} \times (\vec{v}_b \times \vec{B}_{\rm{com}}) 
\end{align}
where $\vec{B}_{\rm{com}} = a^2 \vec{B}_{\rm phys}$ is the comoving magnetic field, $\vec{B}_{\rm phys}$ is the physical field, and $\vec{v_b}$ is the bulk velocity of the plasma. On large comoving scales, $\ell \gg v_b/a H$, the comoving magnetic field remains constant, giving rise to the standard evolution of the physical field, $\vec{B}_{\rm phys} \propto a^{-2}$.  Below this scale, MHD turbulence develops, suppressing both density perturbations and magnetic fields~\cite{Banerjee:2004df, Trivedi:2018ejz, Jedamzik:2018itu, Kim_1996, Subramanian_1998}. The transition scale is

\begin{align} \label{eqn:transition_scale}
    \ell \sim \frac{\left\langle B_{\rm phys}^2 \right\rangle^{1/2}}{a H \sqrt{
    \rho_m}} \approx 0.1 \, {\rm Mpc} \left( \frac{B}{1 \, {\rm nG}} \right),
\end{align}
where $\left\langle ... \right\rangle$ represents an ensemble average. Assuming no helical fields, the PMF power spectrum is given by

\begin{align}
    \left\langle B_i(\vec{k}) B^*_j(\vec{k}') \right\rangle = (2\pi)^3 \delta^3(\vec{k} - \vec{k}') \left( \delta_{ij} - \frac{k_i k_j}{k^2} \right) \frac{P_B(k)}{2},
\end{align}
where $B_i(\vec{k}) \equiv \int d^3 x B(\vec{x}) \exp(i \vec{k} \cdot \vec{x})$ is the Fourier transform of the magnetic field. The power spectrum is typically parametrized as follows:

\begin{align}
    P_B(k) = A k^{n} e^{- k^2 \ell^2},
\end{align}
where $n$ is the model-dependent spectral tilt, and $\ell$ is the transition scale in~\eqref{eqn:transition_scale}. It is assumed that MHD dissiption is the only cutoff in the PMF power spectrum. PMFs generated during inflation result in nearly scale-invariant spectra with $n \approx -3$. PMFs generated after inflation must be causally connected which constraints $n=2$ \cite{Durrer_2003}.
Constraints on the magnetic field are typically quoted in terms of the comoving field strength, $B_{\rm com}$, at some reference scale $\lambda_B$, typically 1 Mpc 

\begin{align}
    B_{\rm com}(\lambda) \equiv \left\langle B^2(\lambda)\right\rangle^{1/2} \propto \left[\displaystyle\int_0^\infty k^2 P_B(k) dk \right]^{1/2} \nonumber \\
    = B(\lambda_B)  \left(\frac{\lambda}{\lambda_B}\right)^{- \frac{n + 3}{2}} \Gamma^{1/2}\left[ (n+3)/2\right],
\end{align}
where $\Gamma(x)$ is the gamma function. The physical RMS magnetic field at comoving scale $\lambda_{\rm com}$ and redshift $z$ can be represented in terms of the comoving magnetic field at reference scale $\lambda_B$ as

\begin{align} \label{eqn:b_redshift}
    B_{\rm phys}(\lambda_{\rm com}, z) = B_{\rm com}(\lambda_B) \left(\frac{\lambda_{\rm com}}{\lambda_B}\right)^{- \frac{n + 3}{2}} (1+z)^2.
\end{align}
Throughout the paper, we do not assume any particular mechanism to generate PMFs. However, in order to avoid the development of turbulence on scales of interest, we assume that the energy injection scale is much smaller than the transition scale in Eq. \eqref{eqn:transition_scale}.
In the following subsection, we summarize current constraints on SM PMFs and show how these bounds can be mapped onto constraints on dark PMFs.




\subsection{Constraints on standard model primordial Magnetic Fields}
Like SM magnetic fields, dark magnetic fields may also exist on large scales. For these magnetic fields to have appreciable effects on dark sector dynamics, they must be seeded by some mechanism in the early Universe and may be enhanced through dynamo effects as the Universe expands. Such mechanisms fall into two main categories: formation of primordial magnetic fields in the early Universe, and late-Universe dynamo amplification of weak seed fields generated by battery mechanisms.

A common mechanism for primordial seed magnetic field generation is the so-called Biermann battery, in which the plasma pressure gradient and temperature gradient are not aligned \cite{1962ApJ...136..615M, 1950ZNatA...5...65B}. In SM plasmas, the misalignment arises from electron-ion drift effects originating from the mass hierarchy between the two charged species. The Biermann battery is inoperative for charged species with equal masses, such as in the dark $U(1)_D$ model under consideration. In this instance, the contributions to the pressure gradient from the positive and negative species cancel. For more details see Appendix \ref{ap:Ohms}. Kinetic plasma instabilities, such as the Weibel instability~\cite{PhysRevLett.2.83, 10.1063/1.1705933} can also generate and amplify magnetic fields in plasmas with anisotropic velocity distributions~\cite{Medvedev_2006,2009ApJ...693.1133L}. In the SM sector, this is typically studied in the context of merging and collapsing galaxies. These mechanisms could be extended to the dark sector as well if significant velocity anisotropies can be created.

Alternatively, more exotic mechanisms have been considered to generate seed primordial magnetic fields. During the inflation, primordial magnetic fields can be generated by explicitly breaking the conformal invariance of the electromagnetic action. References \cite{PhysRevD.37.2743,1992ApJ...391L...1R,PhysRevD.48.2499,PhysRevLett.75.3796,PhysRevD.62.067301, PhysRevD.89.063002, Fujita_2015, KANDUS20111, Subramanian_2010,Subramanian_2016, Martin_2008} consider direct couplings of the electromagnetic fields to other fields that explicitly break the conformal symmetry during inflation. Additionally, models have been suggested that add an $RA_\mu A^\mu$ term to the action but are strongly disfavored since they contain ghosts \cite{Himmetoglu_2009}.

Lastly, phase transitions in the early Universe could lead to the generation of magnetic fields. In the SM sector, electroweak phase transition and the QCD transition are often considered \cite{Brandenburg_1996,Banerjee_2004}. The dark $U(1)_D$ could be the result of some other phase transition in the dark sector or from couplings to SM. First-order phase transitions provide ideal generation mechanisms as colliding bubble walls can seed turbulent fields. Reference \cite{VACHASPATI1991258} also claims that the gradient of the Higgs vacuum expectation value can induce electromagnetic fields at an efficient rate during electroweak phase transition.

After seed field generation, the primordial magnetic fields may be enhanced exponentially by dynamo processes. For turbulent fields, the fluctuation dynamo \cite{1968JETP...26.1031K} provides the most promising growth mechanism. Given we are in a regime of very high magnetic Reynolds number, the seed field can grow rapidly \cite{1986ESASP.251..557K} but may not grow optimally due to the DM's compressible nature \cite{1985ZhETF..88..487K}.


Any magnetic field (SM or dark) will contribute to the Hubble parameter governing the expansion rate of the Universe. If magnetic fields are too large, then the correct abundance of $^4$He will not be produced. Ref. \cite{Grasso_1996} finds that an upper limit of $B_{\rm{phys}}<2\times10^{11}$ Gauss can be placed at $T=10^9$ K for magnetic fields that are uniform over scales larger than the Hubble radius. This corresponds to $B_\text{phys}<0.7\,\mu\text{G}$ today. 





\begin{figure}[tb]
    \centering
    \includegraphics[width=
    0.75\linewidth]{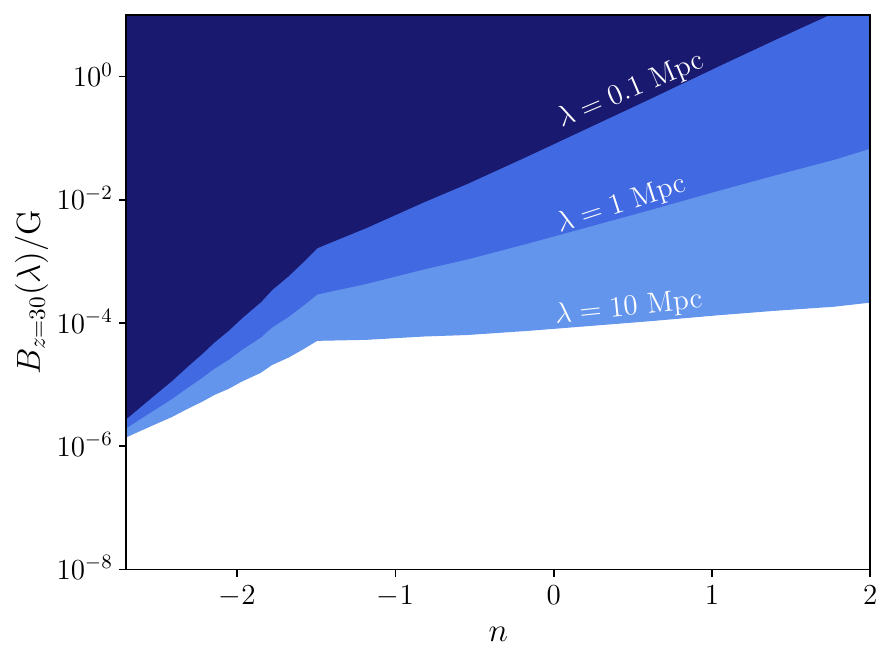}
    \caption{Constraints on the magnetic field strength at different comoving scales: $\lambda = 0.1$ Mpc (dark blue), 1 Mpc (medium blue), and 10 Mpc (light blue) at redshift $z=30$ as a function of spectral tilt of the magnetic power spectrum, $n$, reproduced from Ref.~\cite{PhysRevD.61.043001}.}
    \label{fig:b_lim_z30}
\end{figure}

Random magnetic fields may also generate tensor perturbations leading to anisotropies in the CMB through their coupling to gravitational waves. Ref.~\cite{PhysRevD.61.043001} find that for a scale invariant spectrum on galactic scales, $B_\text{phys}<1$ nG. For a spectral index $n<-3/2$, $B_\lambda<7.9e^{3n}\,\mu\text{G}$ and for a spectral index of $n>-3/2$, $B_\lambda<95e^{0.37n}\,\text{nG}$ where $B_{\lambda}$ is the amplitude at $\lambda=0.1\,h^{-1}\,\rm{Mpc}$. These constraints, translated to constraints at $z=30$ using Eq.~\eqref{eqn:b_redshift}, are shown in Fig.~\ref{fig:b_lim_z30}.

\section{Effects on Matter Power Spectrum} \label{sec_5}

The primordial matter spectrum $P(k)$ is a power spectrum that is used to quantify the fluctuations in the DM density in the Universe following cosmological inflation with respect to the inverse length scale $k$, with units $h/\rm Mpc$ in a comoving reference frame. Alternative DM models beyond the $\Lambda$CDM paradigm can potentially alter the matter power spectrum; thus, in this section, we present the alteration of the matter power spectrum due to the presence of a magnetized dark plasma.

We have demonstrated that through the alteration of the Jeans length in the directions perpendicular to a uniform background dark magnetic field, collective effects in the dark sector as a result of long-range interaction produce an anisotropic pressure support controlled by the dark magnetic field oriented in the direction $\hat{\boldsymbol b}$ with coherence length $L_B$. Linear density modes then obey
\begin{equation}
\ddot\delta_{\boldsymbol k}+2H\dot\delta_{\boldsymbol k}
+\left[\frac{c_s^2(\mu)k^2}{a^2}-4\pi G\bar\rho_m\right]\delta_{\boldsymbol k}=0,
\label{eq:linear-evolution}
\end{equation}
with
\begin{equation}
c_s^2(\mu)=c_{s,\parallel}^2\,\mu^2+c_{s,\perp}^2(1-\mu^2),\qquad
\mu\equiv\hat{\boldsymbol k} \cdot \hat{\boldsymbol b},
\label{eq:lin-eq}
\end{equation}
where $\bar\rho_{m}$ is the average matter density, it is defined $\bar\rho_{m} = \Omega_{m}\rho_{\rm{crit}}$ where $\Omega_{m} \sim 0.3$ today, which implies a directional Jeans scale

\begin{equation}
k_J^2(\mu,a)=\frac{4\pi G\bar\rho_m(a)\,a^2}{c_s^2(\mu)}.
\label{eq:kJ}
\end{equation}

For this section, we set the matter density, $\rho$, equal to $\bar\rho_m$. The derivation in Sec.~\ref{sec:DarkPlasmas} shows that the restoring forces differ for perturbations parallel and perpendicular to the background field direction $\hat{\boldsymbol b}$. 

\begin{figure*}[tb]
\centering
\includegraphics[scale=0.408]{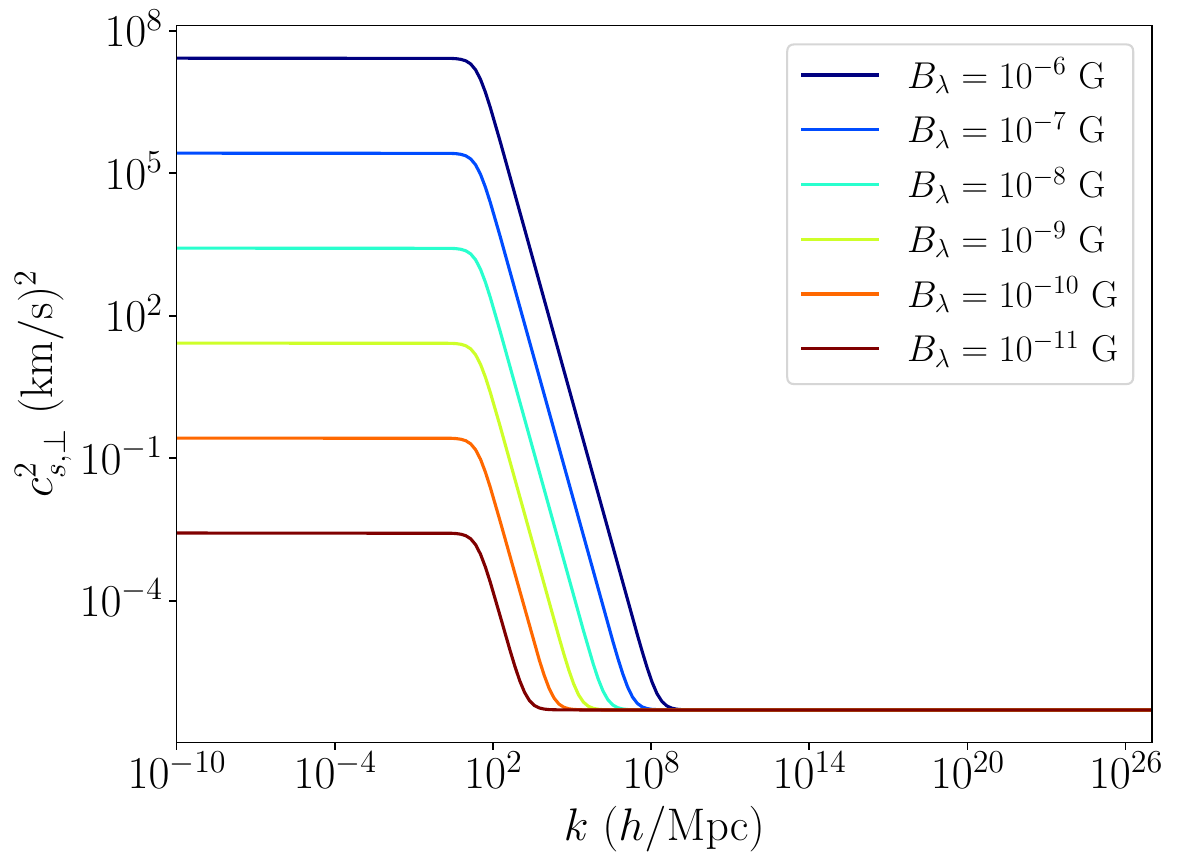} 
\includegraphics[scale=0.408]{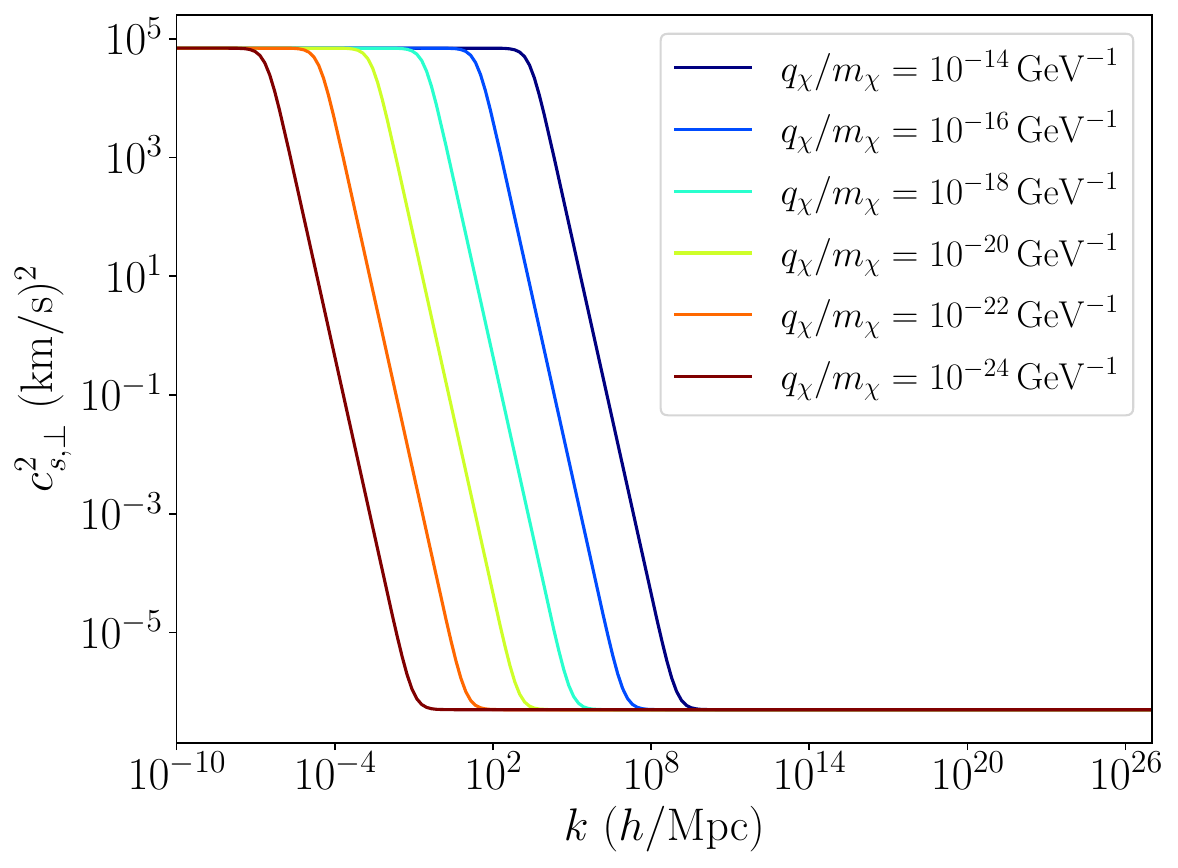} 
\caption{In the plots above, we show the perpendicular group velocity as a function of wave number.  We set our reference scale for the magnetic field spectrum to be $\lambda = 0.1 \, h^{-1} \, \rm Mpc$. Left: the charge-to-mass ratio has been fixed to $10^{-16}\, \rm GeV^{-1}$ and we vary over different strengths of the dark magnetic field. Right: here, the dark magnetic field has been fixed to $B_\lambda = 10^{-6}$ G and we vary over different charge-to-mass ratios.} 
\label{fig:perpendicular_phase_speed_B}
\end{figure*}

While assuming that the coherence length of the dark magnetic field is larger than the MHD turbulence scale defined in Eq.~\eqref{eqn:transition_scale} ($L_B\gg\ell$),
the effective group velocity, $c_{s}^{2}$ from Eq.~\eqref{eq:lin-eq}, acquires a $k$- and angle-dependent pressure term, via Eq.~\eqref{eq:parallel_sound_speed} and Eq.~\eqref{eq:perp_sound_speed}:

\begin{widetext}
\begin{equation}
c_s^2(\mu,k,z)= 3v_{\rm th}^2\,\mu^2
+\left[\frac{1+\beta}{2}v_{\rm th}^2  + \beta\frac{B_{\rm{com}}^2}{\rho(z)}
\left(\frac{k}{k_B}\right)^{n+3}(1+z)^4
\right](1-\mu^2);
\label{eq:cs-final}
\end{equation}
\end{widetext}

We choose fiducial values of $v_{\rm th}$ and $\rho$ to be $\sim 18~\cms$ and $\sim 40\Msol~\kpc^{-3}$ at $z = 0$. In addition, we set the reference scale $\lambda=0.1\,h^{-1}\,\rm{Mpc}$ which translates to $k_{B}=\frac{2\pi}{\lambda}\sim63~ h/\rm{Mpc}$ for the rest of the paper. In Fig.~\ref{fig:perpendicular_phase_speed_B} (left) we plot the perpendicular sound speed for various values of the magnetic field today, $B_\lambda$, with a fixed value of $q_{\chi}/m_{\chi}$ and in Fig.~\ref{fig:perpendicular_phase_speed_B} (right) we plot the perpendicular sound speed for various values of $q_{\chi}/m_{\chi}$ with a fixed value of $B_\lambda$. In both cases, the magnetic field couples to the dark plasma for sufficiently large scales and completely decouples for sufficiently small scales. This can also be seen in the definition of $\beta$ in Eq.~\eqref{eq:chi-equation}, where $0<\beta<1$ for all parameters. We can see that changing the magnetic field alters the sound speed's amplitude, whereas the charge-to-mass ratio determines the transition scale at which the dark plasma couples to the magnetic field and behaves as an ideal MHD fluid.

For the purpose of comparing altered matter power spectra from alternative DM models to that of CDM, it is convenient to define the transfer function, given by
\begin{equation}
    T(k)^2 = \frac{P(k)}{P_{\Lambda\rm{CDM}}(k)},
\end{equation}
where $P(k)$ is the power spectrum under an alternative DM model, and $P_{\Lambda\rm CDM}(k)$ is the isotropic CDM matter power spectrum. Here, we only study the effects of the linear evolution of density perturbations in Eq. \eqref{eq:linear-evolution}. Though we expect our model to additionally impact the nonlinear matter power spectrum, such effects would require dedicated simulation which we consider to be beyond the scope of this paper. In our model, to compare with the isotropic matter power spectrum, we need to integrate over the directionality parameter $\mu$. To do so, we expand the power spectrum in terms of Legendre polynomials, $\mathcal{L}_\ell(\mu)$,

\begin{equation}
    P(k, \mu) = \sum_{\ell =0}P_{\ell}(k)\mathcal{L}_\ell(\mu),
\end{equation} 

\noindent which admits Legendre multipoles (for $\ell=0,2,\dots$),

\begin{equation}
P_\ell(k,z) = \frac{2\ell+1}{2} \int_{-1}^{1} d\mu~ P(k,\mu,z)~ \mathcal{L}_\ell(\mu) .
\label{eq:Pl}
\end{equation}

\begin{figure}[t]
\centering
\includegraphics[width=\linewidth]{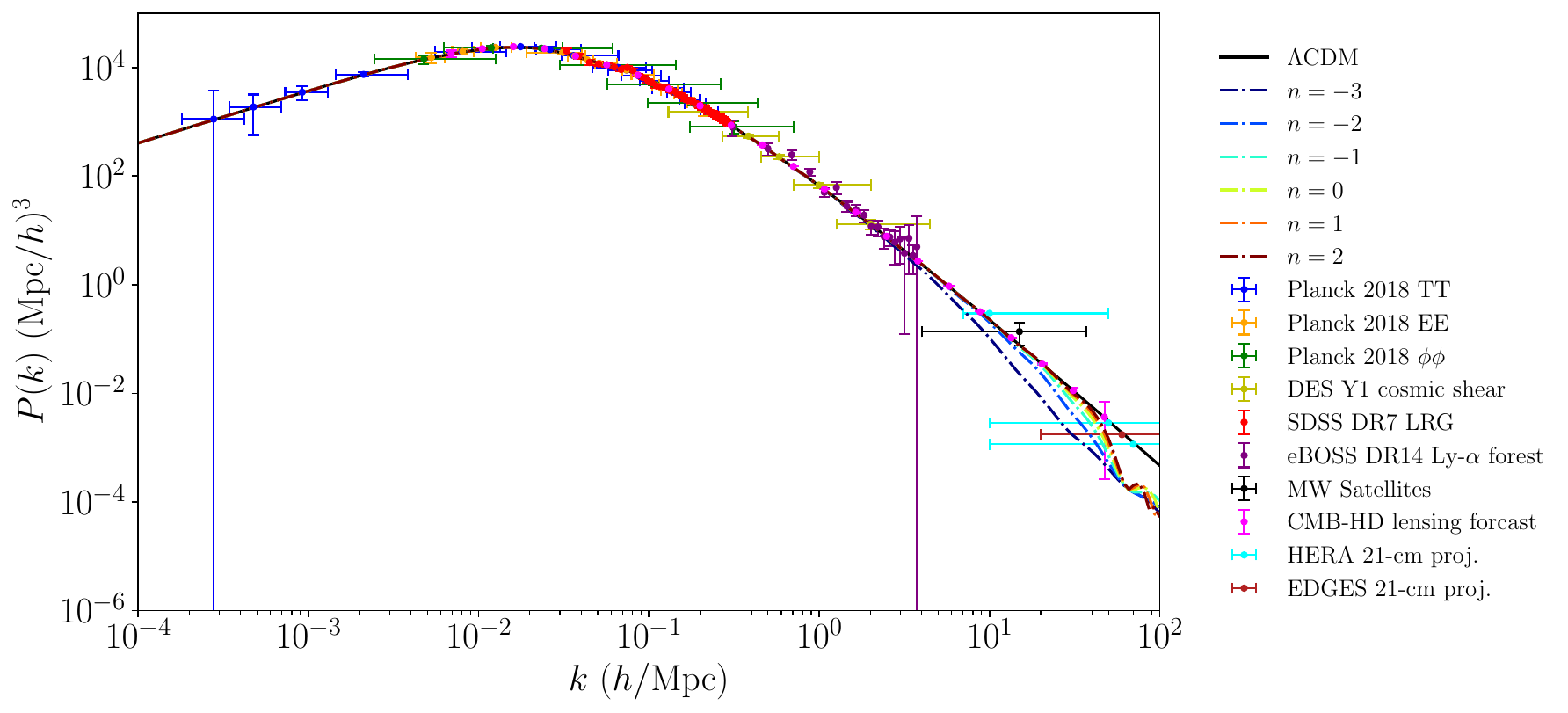}
\caption{Here we show the isotropic power spectrum for a fixed charge-to-mass ratio, $q_\chi/m_\chi = 10^{-18}~\GeV^{-1}$, and a fixed background dark magnetic field value of $B_{\lambda} = 5\times10^{-9}~\rm{Gauss}$ for different spectral indices against the current measurements as well as future/planned surveys (see text for details). We plot using different colored dotted lines to denote various scenarios of a background dark magnetic field, based on constraints derived in the previous section.}
\label{fig:power-spectrum-P0}
\end{figure}

\begin{figure}[b]
\centering
\includegraphics[width=\linewidth]{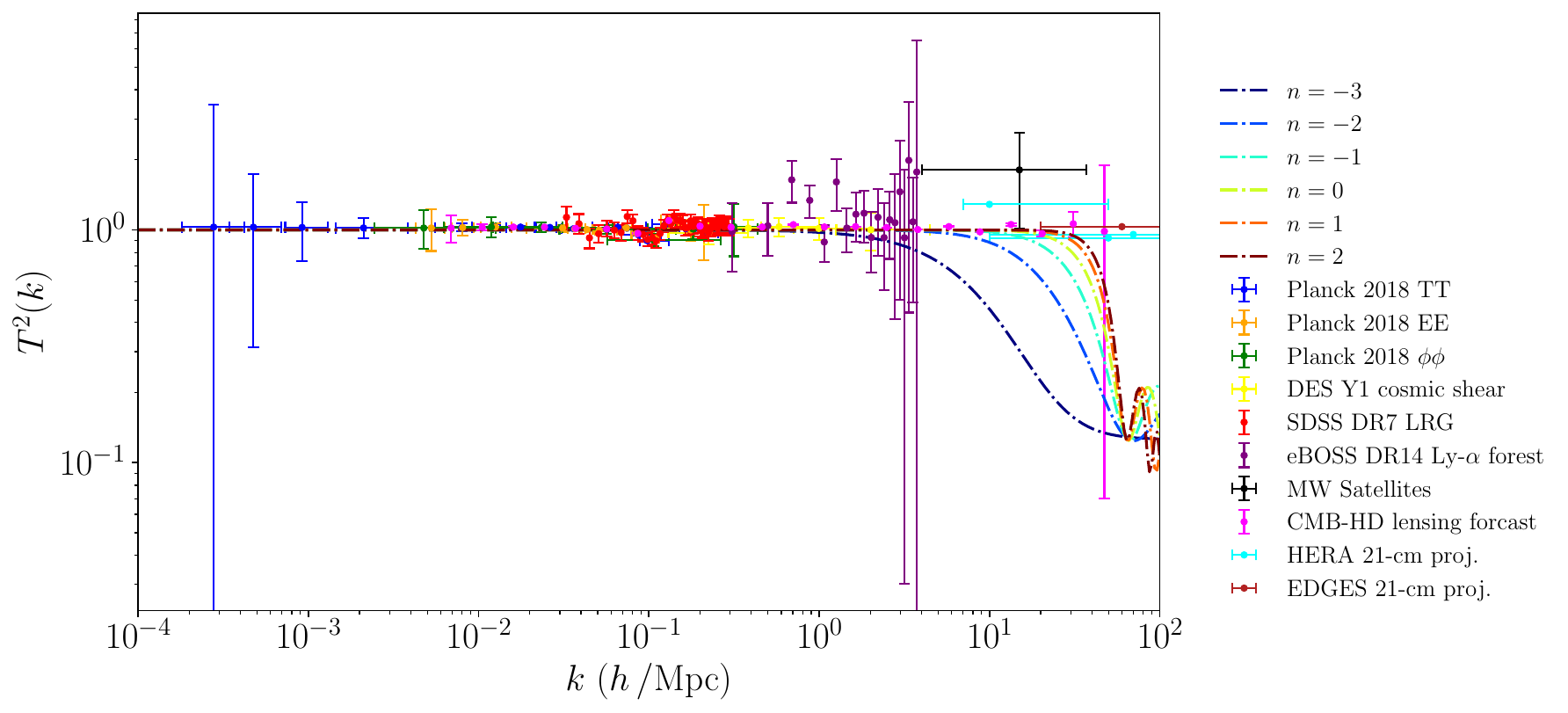}
\caption{Here we show the transfer function at a fixed charge-to-mass ratio, $q_\chi/m_\chi = 10^{-18}~\GeV^{-1}$, and a fixed background dark magnetic field value of $B_{\lambda} = 5\times10^{-9}~\rm{Gauss}$ for different spectral indices. We have overlaid current measurements and future/planned surveys (see text for details).} 
\label{fig:Transfer Function-T0}
\end{figure}

Thus, the $\ell = 0$ multipole will be the isotropic matter power spectrum from our model. In Fig.~\ref{fig:power-spectrum-P0}, we plot several examples of $P_{0}(k,0)$ for various values of the spectral index $n$. We show the current data from Planck 2018~\cite{Planck:2018nkj,Chabanier:2019eai}, DES Y1~\cite{DES_Y1}, SDSS DR7~\cite{SDSS_DR7}, eBOSS DR14~\cite{Abolfathi_2018}, MW Satellites~\cite{MW_forecast}, as well as projected future measurements from CMB-HD lensing~\cite{MacInnis_2025}, and the HERA and EDGES 21-cm projections~\cite{HERA,EDGES,Munoz:2019hjh}. In Fig.~\ref{fig:Transfer Function-T0} we plot the corresponding transfer function of each of the curves from Fig.~\ref{fig:power-spectrum-P0}. We note that, unlike warm DM or SIDM models, we do not see pure exponential suppression or dark acoustic oscillations (DAOs), but rather a smooth reduction/alteration in power, followed by a continued power-law fall-off, with DAOs occurring at much larger wave numbers. In principle, there is a scale such that all power is suppressed, though we do not probe these scales. Additionally, it has recently been shown that the solutions to Eq.~\eqref{eq:lin-eq} closely match those from state-of-the-art numerical MHD simulations for scales larger than the MHD turbulent scale, $\lambda\gg\ell$~\cite{Ralegankar_2025}.

To place constraints on magnetic fields in the secluded $U(1)_D$ gauge sector, we scan over a 2D parameter space of the magnetic field today at reference scale $\lambda=0.1\,h^{-1}\,\rm{Mpc}$, and the DM charge-to-mass ratio. At each point in the scan, we numerically solve Eq.~\eqref{eq:lin-eq} to generate $P_0(k,0)$ and calculate the $\chi^2$ statistic against the current and future power spectrum data given in Fig.~\ref{fig:power-spectrum-P0}. The resulting parameters that can be excluded at the 68\%, 95\%, and 99.73\% confidence levels are shown in the solid pink, purple, and red colored regions, respectively. Along with these contours, we also show projected constraints from future observations given by the dashed lines---the color scheme has the same meaning.

In each plot, we also fix the spectral index, $n$, to several representative values. For each index, the region is constrained from above by limits from CMB anisotropies~\cite{PhysRevD.61.043001}, noting that the strongest constraints are for the scale invariant power spectrum, $n=-3$. Simulations of dark plasma instabilities in the Bullet Cluster additionally constrain the DM charge-to-mass ratio $q_\chi/m_\chi<2\times10^{-14}\,\rm{GeV}^{-1}$~\cite{Giffin}. Lastly, the weak gravity conjecture~\cite{Arkani_Hamed_2007} mandates that theories of quantum gravity with a $U(1)_D$ gauge field must satisfy $q_\chi/m_\chi>1/M_P$, where $M_P$ is the reduced Planck mass. This additionally constrains the dark charge-to-mass ratio from below. We find that with current data, the dark magnetic fields that have not already been excluded by measurements of the CMB anisotropies (with the small exception of the $n=-2$ case, see Fig.~\ref{fig:parameter-space-constraints}) do not significantly alter the linear matter power spectrum, and are therefore consistent with current cosmology. With future measurements at higher wave numbers, we expect to be able to probe deviations from the CDM power spectrum.

\section{Discussion and Conclusions} \label{sec_6}

\begin{figure*}[h]
\centering
\includegraphics[scale=0.90]{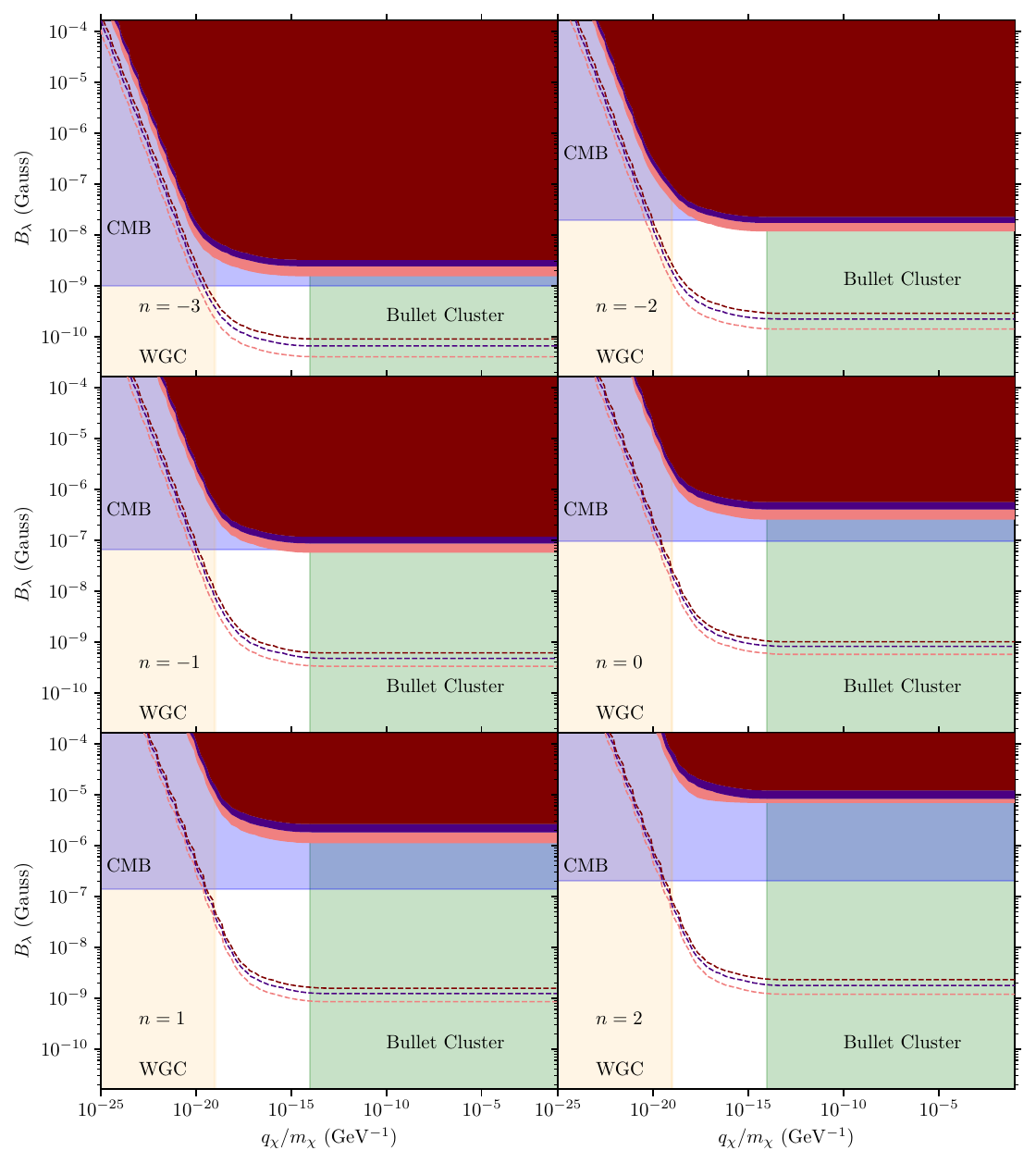}
\caption{ Here we show the resultant parameter space for scans using our effective group velocity Eq.~\eqref{eq:cs-final} to grow the linear density modes via Eq.~\eqref{eq:lin-eq} for current (shaded regions) and future (dashed lines) measurements of the linear matter power spectrum shown in Fig.~\ref{fig:power-spectrum-P0} (see text for the data details). The region shown in pink is excluded at the $1\sigma$ level, the purple is excluded at the $2\sigma$ level, and the red region is excluded by a minimum of $3\sigma$. We also show constraint regions from the CMB~\cite{PhysRevD.61.043001} (blue), Bullet Cluster~\cite{Giffin} (green), and the weak gravity conjecture (yellow)~\cite{Arkani_Hamed_2007}.}
\label{fig:parameter-space-constraints}
\end{figure*}

In this work, we show that in the presence of large-scale dark magnetic fields, DM halos charged under a secluded $U(1)_D$ gauge field experience magnetic pressure that delays gravitational collapse via the Jeans instability. For correlation lengths on scales much larger than the turbulent scale, $\ell$, we solve Eq.~\eqref{eq:lin-eq} to map the effects onto the linear matter power spectrum. Figure \ref{fig:parameter-space-constraints} shows that current measurements of the linear matter power spectrum provide constraints weaker than constraints set by tensor modes of the CMB. However, we expect future observations to probe into unconstrained regions of parameter space, further closing the gap between the constraint set by Ref.~\cite{Giffin} and the weak gravity conjecture.

This work can be extended in several new directions to incorporate well-motivated models of DM. Kinetic mixing between the SM photon and the dark photon can induce a small effective electric charge for particles in the dark sector. As a result, such particles, commonly referred to as millicharged DM, acquire suppressed couplings to SM electromagnetic fields in addition to their interactions with dark sector forces. This scenario has garnered significant interest due to the potential for millicharged DM to interact with both visible and dark sector electromagnetic fields.

Previous literature has examined the interactions of millicharged DM with SM particles and electromagnetic fields. In particular, the evolution of millicharged DM in the presence of plasma instabilities has been used as a powerful tool to constrain its properties, as demonstrated in studies of supernova remnants \cite{Li_2020} and the Bullet Cluster \cite{Cruz_2023}. Considering a similar framework in which millicharged DM couples solely to SM electromagnetic fields, the model used in this work can be extended to incorporate baryonic effects by utilizing a two-fluid MHD model that includes gravity. Though SM primordial fields are generally better understood in the literature, the two-fluid model is sufficiently complex that the system likely needs to be solved numerically.  We intend to consider this scenario in future work.

Another future direction is the consideration of the evolution of anisotropic collapse and halo triaxiality due to dark plasma effects. As a result of the directionally dependent alteration to the Jeans length, halo triaxiality and its radial evolution can serve as a direct macroscopic tracer of microscopic dark sector plasma effects. The anisotropic screening mechanism developed here predicts coherent, orientation-dependent modifications to collapse along and across the local dark-field direction $\hat{\boldsymbol{b}}$, leading to systematic biases in intrinsic shape distributions.  Relative to the CDM baseline, dark-plasma screening enhances the contrast between inner and outer halo shapes and induces characteristic axis twists that preserve memory of the underlying field geometry.  This ``radial memory'' manifests as steeper inner-outer ellipticity gradients and alignment signatures that cannot be replicated by baryonic condensation or isotropic self-interaction models.  Triaxiality therefore emerges as a sensitive diagnostic of anisotropic support in the dark sector, complementing traditional small-scale power-spectrum constraints.

Several independent probes can access these shape signatures across mass and redshift.  Weak-lensing quadrupoles and projected ellipticity profiles offer stackable, statistical tests of enhanced inner–outer contrast; strong-lensing reconstructions at  $r\!\sim\!0.05$–$0.2\,R_{\mathrm{vir}}$ provide direct measures of central isopotential flattening and axis twists; and x-ray or SZ isophote ellipticities trace the same anisotropy in the intracluster gas.  Correlations between halo orientations and large-scale filaments, as well as anisotropies in satellite and splashback distributions, further constrain the coherence and preferred direction of $\hat{\boldsymbol{b}}$.  

In Milky Way–like and more massive disk galaxies, however, a strong stellar bar can imprint nonaxisymmetric signatures on both the luminous and dark components, including boxy/peanut bulges, isophotal twists, and apparent inner–outer misalignments \cite{Seidel:2015BaLROG,DiazGarcia:2016Bars,Kim:2016Bars,Collier:2021HaloBars,Marostica:2024HaloBar}.  These bar-driven distortions are largely confined to radii $\lesssim(1$–$2)\,R_{\rm d}$ and are tied to the presence, pattern speed, and orientation of the stellar disk, in contrast with the large-scale, approximately radius-independent alignment predicted for the anisotropic-screening axis $\hat{\boldsymbol{b}}$.  At dwarf and low-mass-galaxy scales, hydrodynamic simulations and H\,I kinematic studies show that baryons can produce both oblate and mildly prolate dark-matter haloes, with the halo flattening and triaxiality correlating with gas fraction and the development of a stellar disk \cite{Das:2023Oblateness,Das:2023GasFrac,Orkney:2023EDGE,Keith:2025Marvel}.  Gas-rich dwarfs with $M_{\rm gas}/M_{\rm baryon}\gtrsim 0.5$ tend to host oblate haloes, whereas more stellar-dominated systems span a wide range of shapes from significantly flattened to nearly spherical or slightly prolate \cite{Das:2023Oblateness,Das:2023GasFrac}.  In $\Lambda$CDM, baryonic condensation in more massive dwarfs and spirals also tends to round the central halo relative to dark-matter–only expectations \cite{Zavala:2019Review}, while ultrafaint, gas-poor dwarfs retain the strongly prolate shapes of collisionless haloes \cite{Orkney:2023EDGE}.  Our plasma-mediated scenario instead predicts a coherent preferred axis for the inner halo that can persist even in gas-poor systems, so a joint detection of (i) an unusually steep ellipticity gradient or twist relative to $\Lambda$CDM expectations and (ii) a consistent alignment axis across dwarfs, massive disks (with and without bars), and clusters would be difficult to reproduce with purely baryonic or elastic-SIDM rounding alone \cite{Peter:2013SIDMShapes}.  Such cross-comparisons would establish halo triaxiality as a powerful, multiobservable avenue for testing plasma-mediated structure formation in the dark sector.

\section{Acknowledgments}
The research of P.G., S.P., and M.G.R. is supported, in part, by the U.S. Department of Energy (DOE) Grant No. DE-SC0010107. The research of P.G. is additionally supported in part by the Achievement Rewards for College Scientists Foundation (ARCS) 2025-2026. A.P. acknowledges support from the Princeton Center for Theoretical Science and from the DOE under Award Number DE-SC0007968.

\bibliography{new_ref}

\clearpage
\newpage
\appendix 

\section{DERIVATION OF GENERALIZED OHM'S LAW}
\label{ap:Ohms}
We now aim to justify our choice of closure of the typical MHD system of equations through the generalized Ohm's law. The equations of motion of the positive and negative species of particles can be described by the following differential equations
\begin{align}
    &\partial_t(\rho_+\vec{U}_+)+\nabla\cdot(\rho_+\vec{U}_+\vec{U}_+)=
    \rho_+\frac{q_\chi}{m_\chi}\left(\vec{E}+\vec{U}_+\times\vec{B}\right)\nonumber\\
    &\qquad\qquad\qquad\qquad-\nabla\cdot\stackrel{\leftrightarrow}{P}_+-\rho_+\nu_{\pm}(\vec{U}_+-\vec{U_-}),
\end{align}
\vspace{1cm}
\begin{align}
    &\partial_t(\rho_-\vec{U}_-)+\nabla\cdot(\rho_-\vec{U}_-
        \vec{U}_-)=
    -\rho_-\frac{q_\chi}{m_\chi}\left(\vec{E}+\vec{U}_-\times\vec{B}\right)\nonumber\\
    &\qquad\qquad\qquad\qquad-\nabla\cdot\stackrel{\leftrightarrow}{P}_--\rho_-\nu_{\pm}(\vec{U}_--\vec{U_+}),
\end{align}
where $\rho_{+(-)}$ is the mass density of the positive (negative) species, $\vec{U}_{+(-)}$ is the average velocity of the positive (negative) species, and $\nu_\pm$ is the frequency of Coulomb scatterings between the positive and negative species. Taking the difference between these two equations we arrive at the generalized Ohm's law for pair plasmas.
\begin{widetext}
\begin{equation}
    \label{eq:GenOhm}
    \vec{E}+\vec{U}\times\vec{B}=\eta\vec{J}+\frac12\frac{m_\chi^2}{q_\chi^2\rho}\left[\partial_t\vec{J}+\nabla\cdot(\vec{U}\vec{J}+\vec{J}\vec{U})\right]+\frac{m_\chi}{q_\chi\rho}\nabla\cdot\left(\stackrel{\leftrightarrow}{P}_+-\stackrel{\leftrightarrow}{P}_-\right)
\end{equation}
\end{widetext}
where we have now defined the fluid mass density $\rho=\rho_++\rho_-$, the fluid bulk velocity $\vec{U}=(\vec{U}_++\vec{U}_-)/2$, and the current $\vec{J}=q_\chi\rho(\vec{U}_+-\vec{U}_-)/m_\chi$ as well as the resistivity of the plasma
\begin{equation}
    \eta=\frac{m_\chi^2\nu_\pm}{q_\chi^2\rho}.
\end{equation}
We now consider the scenario under study. Given that we expect the positively changed species to be in thermal equilibrium with the negatively charged species, $\stackrel{\leftrightarrow}{P}_- =\stackrel{\leftrightarrow}{P}_+$ and the last term in Eq. (\ref{eq:GenOhm}) vanishes. Next, we assume that the unperturbed plasma initially has no net current or bulk velocity. Hence, after linearizing Eq. (\ref{eq:GenOhm}), $\nabla\cdot(\vec{U}\vec{J}+\vec{J}\vec{U})$ will only contain second order terms which we neglect in this work. Lastly, due to the strong constraints already placed on the dark charge-to-mass ratio by \cite{Giffin}, we expect $\nu_\pm\ll T^{-1}$, where $T$ is the timescale of gravitational collapse of the plasma. Hence, the resistive term of Eq. (\ref{eq:GenOhm}) is negligible compared to the $\partial_t\vec{J}$ term. Thus, we chose to implement the following Ohm's law for our study
\begin{equation}
    \vec{E}+\vec{U}\times\vec{B}=\frac12\frac{m_\chi^2}{q_\chi^2\rho}\partial_t\vec{J}
\end{equation}
Note that in the limit of $q_\chi/m_\chi\to\infty$ we obtain the Ohms law for an ideal MHD plasma
\begin{equation}
    \vec{E}+\vec{U}\times\vec{B}=0.
\end{equation}

\section{DERIVATION OF DISPERSION RELATION}
\label{ap:Disp}
In this Appendix, we explicitly show the full derivation of the dispersion relation in Eq. (\ref{eq:Disp_Rel}). These calculations follow very similarly to those performed in Ref. \cite{Gliddon_1966}. However, we consider here effects from nonideal MHD conditions. In the presence of an external magnetic field, a self-gravitating plasma obeys the following system equations
\begin{align}
    &\nabla\cdot \vec{E} = \rho_c & \nabla \times \vec{E} = -\frac{\partial \vec{B}}{\partial t}\\
    & \nabla \cdot \vec{B} = 0 & \nabla \times \vec{B} =  \vec{J}+ \frac{\partial\vec{E}}{\partial t}\\
    &\frac{\partial \rho_m}{\partial t}+\nabla \cdot (\rho_m \vec{U})=0 & \rho_m \frac{d\vec{U}}{dt}=\vec{J}\times \vec{B} -\nabla \stackrel{\leftrightarrow}{P}-\rho_m\nabla V\\
    & \frac{d}{dt}\left(\frac{p_\perp B^2}{\rho_m^3}\right)=0 & \frac{d}{dt}\left(\frac{p_\parallel}{\rho_m B}\right)=0\\
    &\vec{E}+\vec{U}\times\vec{B}=\frac12\frac{m_\chi^2}{q_\chi^2\rho_m}\frac{\partial\vec J}{\partial t} & \nabla^2V=4\pi G\rho_m
\end{align}
Here we have utilized our choice of Ohm's Law discussed in Appendix \ref{ap:Ohms} as well as the Chew-Goldberger-Low equation of state for cylindrically symmetric magnetized plasmas \cite{CGL:1956}. 
As in typical MHD studies, we assume that the plasma remains net neutral during its evolution and neglect effects due to separation of charges by setting $\rho_c=0$. Additionally, we assume that the system is varying slowly enough in time such that displacement current $\partial\vec{E}/\partial t$, can be neglected when compared to $\vec{J}$. This allows $\vec{J}$ to be eliminated from the system of equations
\begin{align}
    &\nabla\cdot \vec{E} = 0, \qquad \nabla \times \vec{E} = -\frac{\partial \vec{B}}{\partial t},\\
    & \nabla \cdot \vec{B} = 0,  \qquad \frac{\partial \rho}{\partial t}+\nabla \cdot (\rho \vec{U})=0, \\
    &\rho \frac{d\vec{U}}{dt}=(\nabla\times\vec{B})\times \vec{B} -\nabla \stackrel{\leftrightarrow}{P}-\rho\nabla V,\\
    & \frac{d}{dt}\left(\frac{p_\perp B^2}{\rho^3}\right)=0, \qquad \frac{d}{dt}\left(\frac{p_\parallel}{\rho B}\right)=0,\\
    &\vec{E}+\vec{U}\times\vec{B}=\frac12\frac{m_\chi^2}{q_\chi^2\rho}\frac{\partial}{\partial t}\nabla\times\vec{B}, \qquad \nabla^2V=4\pi G\rho,
\end{align}

Here we have defined $\rho=\rho_m$ for ease of notation. It is important to note that $\nabla\cdot \vec{E} = 0$ and $\nabla \cdot \vec{B} = 0$ only need to be satisfied in the initial conditions. Ensuring the remaining Maxwell's equations are satisfied during the time evolution of the plasma ensures that the initial conditions remain satisfied. As a homogeneous plasma in a homogeneous magnetic field satisfies these conditions, we no longer require them. Further eliminating the electric field from the system yields
\begin{align}
    & \nabla \times \left(\vec{U}\times\vec{B}-\frac12\frac{m_\chi^2}{q_\chi^2\rho}\frac{\partial}{\partial t}\nabla\times\vec{B}\right) = \frac{\partial \vec{B}}{\partial t}\\ 
    & \nabla^2V=4\pi G\rho, \qquad \frac{\partial \rho}{\partial t}+\nabla \cdot (\rho \vec{U})=0, \\
    &\rho \frac{d\vec{U}}{dt}=(\nabla\times\vec{B})\times \vec{B} -\nabla \stackrel{\leftrightarrow}{P}-\rho\nabla V,\\
    & \frac{d}{dt}\left(\frac{p_\perp B^2}{\rho^3}\right)=0, \qquad \frac{d}{dt}\left(\frac{p_\parallel}{\rho B}\right)=0.
\end{align}
Note that in the $q_\chi/m_\chi  \to\infty$ limit, we obtain the same system of equations studied under ideal MHD conditions in Ref. \cite{Gliddon_1966}. 

We now define the system which we wish to study. Here, we consider an infinitely large homogeneous plasma with a uniform magnetic field oriented in the $z$ direction. We define
\begin{equation}
    \vec{B}=(0,0,B_0)
\end{equation}
and a pressure tensor
\begin{equation}
   \stackrel{\leftrightarrow}{P}=
    \begin{bmatrix}
        p_\perp & 0 & 0\\
        0 & p_\perp & 0\\
        0 & 0 & p_\parallel
    \end{bmatrix}
    =p_\perp I+(p_\parallel-p_\perp)\hat{z}\hat{z}.
\end{equation}
Initially, the plasma is at rest with $\vec U=0$ with some finite mass density $\rho=\rho_0$. We consider the plasma to initially have isotropic pressure $p_\perp=p_{0}$ and $p_{ \parallel}=p_{ 0}$. However, the plasma pressure may evolve anisotropically in directions parallel or perpendicular to the background magnetic field.

We now proceed to linearize the system, considering small perturbations we replace
\begin{align}
&\rho\to \rho_0+\delta\rho,\quad \vec{U}\to \delta \vec{U},\quad \stackrel{\leftrightarrow}P\to \stackrel{\leftrightarrow}P_0+\stackrel{\leftrightarrow}{\delta P}, \\
&\quad \vec B\to \vec B_0+\delta \vec B, \quad V\to \delta V_0.
\end{align}
Keeping only first order terms, we obtain the new system of equations
\begin{equation}
    \label{eq:Cons_Mass}
    \frac{\partial}{\partial t}\delta \rho+\rho_0\nabla\cdot \delta \vec U=0
\end{equation}
\begin{equation}
    \label{eq:Cons_Mom}
    \rho_0\frac{\partial}{\partial t}\delta \vec U=-\nabla\cdot \stackrel{\leftrightarrow}{\delta P}+(\nabla\times\delta \vec B)\times \vec B_0-\rho_0\nabla \delta V
\end{equation}
\begin{equation}
    \label{eq:AD1}
    \frac{\delta p_\parallel}{p_{ 0}}+\frac{2\delta B}{B_0}=\frac{3\delta \rho}{\rho_0}
\end{equation}
\begin{equation}
    \label{eq:AD2}
    \frac{\delta p_\perp}{p_{0}}=\frac{\delta\rho}{\rho_0}+\frac{\delta B}{B_0}
\end{equation}
\begin{equation}
    \label{eq:MHD_1}
    \frac{\partial}{\partial t}\delta \vec B=-\frac12\frac{m_\chi^2}{q_\chi^2\rho}\frac{\partial}{\partial t}\nabla\times(\nabla\times\delta \vec B)+\nabla\times(\delta \vec U \times \vec B_0)
\end{equation}
\begin{equation}
    \label{eq:Grav}
    \nabla^2 \delta V = 4\pi G \delta \rho
\end{equation}
where we have defined the scalar quantity $ \delta B=\vec{B}_0\cdot\delta\vec{B}/|\vec{B_0}|$. We now assume all perturbed quantities are proportional to $\mathrm{exp}\, i(\Vec{k}\cdot\Vec{r}-\omega t)$ and replace derivative operators. Additionally, we substitute $\delta U$ for the time derivative of a Lagrangian displacement of the fluid, $\xi$
\begin{equation}
    \delta \vec U=\frac{\partial \vec\xi}{\partial t}=-i\omega\vec\xi
\end{equation}  

Beginning with Eq. (\ref{eq:MHD_1})
\begin{equation}
    -i\omega  \delta\Vec{B}=i\vec{k}\times\left(-i\Vec{k}\times(-i\omega) \frac12\frac{m_\chi^2}{q_\chi^2\rho}\delta\vec{B}-i\omega\Vec{\xi}\times B_0\hat{z}\right)
\end{equation}
yields the solution
\begin{equation}
    \delta\Vec{B}=iB_0\beta(k_z\xi_x\hat{x}+k_z\xi_y\hat{y}-(k_x\xi_x+k_y\xi_y)\hat{z}).
\end{equation}
where
\begin{equation}
    \beta=\frac{1}{1+\frac{k^2}{2\rho}\frac{m_\chi^2}{q_\chi^2}}
\end{equation}
which implies the following form for the scalar $\delta B$,
\begin{equation}
    \delta B=\frac{\vec{B}_0\cdot\delta\vec{B}}{B_0}=-iB_0\beta(k_x \xi_x+k_y\xi_y)
\end{equation}
Next, from Eq. (\ref{eq:Cons_Mass})
\begin{equation}
    \frac{\delta\rho}{\rho_0}=-i\vec{k}\cdot\vec{\xi}
\end{equation}
Can be utilized along with Eqs. (\ref{eq:AD1}) and (\ref{eq:AD2}) to obtain
\begin{equation}
    \frac{\delta p_\parallel}{p_{0}}=-i(3-2\beta)(k_x\xi_x+k_y\xi_y)-3i(k_z\xi_z)
\end{equation}
and 
\begin{equation}
    \frac{\delta p_\perp}{p_{0}}=-i(1+\beta)(k_x\xi_x+k_y\xi_y)-i(k_z\xi_z)
\end{equation}
Next, we first start by determining $\stackrel{\leftrightarrow}{\delta P}$
\begin{equation}
    \stackrel{\leftrightarrow}{\delta P}=\delta p_\perp I+(\delta p_\parallel-\delta p_\perp)\hat{z}\hat{z}
\end{equation}

We then find the $x$, $y$, and $z$ components of $\nabla\cdot\stackrel{\leftrightarrow}{\delta P}$ to be
\begin{equation}
    (i\vec{k}\cdot \stackrel{\leftrightarrow}{\delta P})_x=\left[(1+\beta)(k_x\xi_x+k_y\xi_y)+k_z\xi_z\right]k_xp_{0}
\end{equation}
\begin{equation}
    (i\vec{k}\cdot \stackrel{\leftrightarrow}{\delta P})_y=\left[(1+\beta)(k_x\xi_x+k_y\xi_y)+k_z\xi_z\right]k_yp_{0}
\end{equation}
\begin{equation}
   (i\vec{k}\cdot \stackrel{\leftrightarrow}{\delta P})_z=[(3-2\beta)(k_x\xi_x+k_y\xi_y)+3(k_z\xi_z)]k_zp_0
\end{equation}
For the next term in Eq. (\ref{eq:Cons_Mom} we have
\begin{align}
(\nabla\times\delta\vec{B})\times\vec{B}_0=&-B_0^2\beta[(k_x(k_x\xi_x+k_y\xi_y)+k_z^2\xi_x)\hat{x}\nonumber\\
&+(k_y(k_x\xi_x+k_y\xi_y)+k_z^2\xi_y)\hat{y}]
\end{align}
From Eq. (\ref{eq:Grav}) we find
\begin{equation}
    \delta V=i\frac{4\pi G \rho_0}{k^2}(\vec{k}\cdot\vec{\xi})
\end{equation}
Together, this yields the following equations for each component of Eq. (\ref{eq:Cons_Mom})
\begin{align}
    \label{eq:Disp1}
    -\rho_0\omega^2\xi_x=&-\left[(1+\beta)(k_x\xi_x+k_y\xi_y)+k_z\xi_z\right]k_xp_{0}\nonumber\\
    &-B_0^2\beta(k_x(k_x\xi_x+k_y\xi_y)+k_z^2\xi_x)\nonumber\\
    &+4\pi G\rho_0^2\frac{k_x}{k^2}(k_x\xi_x+k_y\xi_y+k_z\xi_z)
\end{align}
\begin{align}
    \label{eq:Disp2}
    -\rho_0\omega^2\xi_y=&-\left[(1+\beta)(k_x\xi_x+k_y\xi_y)+k_z\xi_z\right]k_yp_{0}\nonumber\\
    &-B_0^2\beta(k_y(k_x\xi_x+k_y\xi_y)+k_z^2\xi_y)\nonumber\\
    &+4\pi G\rho_0^2\frac{k_y}{k^2}(k_x\xi_x+k_y\xi_y+k_z\xi_z)
\end{align}
\begin{align}
    \label{eq:Disp3}
    -\rho_0\omega^2\xi_z=&-[(3-2\beta)(k_x\xi_x+k_y\xi_y)+3(k_z\xi_z)]k_zp_0\nonumber\\
   &+4\pi G\rho_0^2\frac{k_z}{k^2}(k_x\xi_x+k_y\xi_y+k_z\xi_z)
\end{align}
Having rid of all of the perturbed quantities, we now drop the 0 subscript for ease of notation. We also define

\begin{align}
    &\xi_\parallel=\xi_z,\quad \xi_\perp=(\xi_x,\xi_y),\quad
    k_\parallel=k_z,\\
    &\quad k_\perp=(k_x,k_y),\quad k_\perp^2=k_x^2+k_y^2.
\end{align}
We now multiply Eq. (\ref{eq:Disp1}) by $k_x$ and Eq. (\ref{eq:Disp2}) by $k_y$ and add the resulting equations together to yield
\begin{align}
    &\left[\rho\omega^2-(1+\beta)k_\perp^2p  -\beta B^2k^2+4\pi G\rho^2\frac{k_\perp^2}{k^2}\right]k_\perp\cdot\xi_\perp\nonumber\\
    &\qquad\qquad\qquad\qquad=\left( k_\perp^2 p-4\pi G \rho^2\frac{k_\perp^2}{k^2}\right)k_\parallel\xi_\parallel
\end{align}
Equation (\ref{eq:Disp3}) yields
\begin{align}
    &\left(\rho\omega^2-3k_\parallel^2p+4\pi G \rho^2\frac{k_\parallel^2}{k^2}\right)k_\parallel\xi_\parallel\\
    &\qquad=\left((3-2\chi)k_\parallel^2 p-4\pi G\rho^2 \frac{k_\parallel^2}{k^2}\right)k_\perp\cdot\xi_\perp
\end{align}

Combining the two above results gives the dispersion relation
\begin{widetext}
\begin{multline}
     \left[\rho\omega^2-(1+\beta)k_\perp^2p-\beta B^2k^2+4\pi G\rho^2\frac{k_\perp^2}{k^2}\right]
    \left(\rho\omega^2-3k_\parallel^2p+4\pi G \rho^2\frac{k_\parallel^2}{k^2}\right)\\=k_\perp^2k_\parallel^2\left[\left(p-\frac{4\pi G \rho^2}{k^2}\right)^2+2\left(p-\frac{4\pi G \rho^2}{k^2}\right)\left(1-\beta\right)p\right]
\end{multline}
\end{widetext}

\end{document}